%% file: Template.tex
\documentclass{article}
\usepackage{spconf,amsmath,graphicx}
\usepackage{amsfonts}
\usepackage{minitoc}
\usepackage[acronym,nomain,nonumberlist]{glossaries}
\usepackage{hyperref}
\input{glossary}

\usepackage{float}
\usepackage{stfloats}
\usepackage{multirow}
\usepackage{rotating}
\usepackage{booktabs}
\usepackage{url}

\title{Double Multi-Head Attention for Speaker Verification}
\name{Miquel India, Pooyan Safari, Javier Hernando\thanks{This work was supported by the Spanish project PID2019-107579RB-I00 / AEI / 10.13039/501100011033.}}
\address{TALP Research Center, Department of Signal Theory and Communications,\\
Universitat Politecnica de Catalunya, Barcelona, Spain\\
\{miquel.angel.india, javier.hernando\}@upc.edu,  pooyan.safari@tsc.upc.edu}

\begin{document}
\ninept

\maketitle

\begin{abstract}
Most state-of-the-art Deep Learning systems for speaker verification are based on speaker embedding extractors. These architectures are commonly composed of a feature extractor front-end together with a pooling layer to encode variable-length utterances into fixed-length speaker vectors. In this paper we present Double Multi-Head Attention pooling, which extends our previous approach based on Self Multi-Head Attention. An additional self attention layer is added to the pooling layer that summarizes the context vectors produced by Multi-Head Attention into a unique speaker representation. This method enhances the pooling mechanism by giving weights to the information captured for each head and it results in creating more discriminative speaker embeddings. We have evaluated our approach with the VoxCeleb2 dataset. Our results show $6.09\%$ and $5.23\%$ relative improvement in terms of EER compared to Self Attention pooling and Self Multi-Head Attention, respectively. According to the obtained results, Double Multi-Head Attention has shown to be an excellent approach to efficiently select the most relevant features captured by the CNN-based front-ends from the speech signal.
\end{abstract}

\begin{keywords}
self multi-head attention, double attention, speaker recognition, speaker verification
\end{keywords}

\section{Introduction}
\label{sec:intro}

Speaker verification aims to determine whether a pair of audios corresponds to the same speaker. Given speech signals, speaker verification systems are able to extract speaker identity patterns from the characteristics of the voice. These patterns can be both statistically modelled or encoded into discriminative speaker representations. Over the last few years, researchers have put huge effort on encoding these speaker characteristics into more discriminative speaker vectors. Current state-of-the-art speaker verification systems are based on \gls{dl} approaches. These architectures are commonly trained as speaker classifiers in order to be used as speaker embedding extractors. Speaker embeddings are fixed-length vectors extracted from some of the last layers of these \glspl{dnn} \cite{ghahabi2020deep}. The most known representation is the x-vector \cite{snyder2018x}, which has become state-of-the-art for speaker recognition and has also been used for other tasks such as language and emotion recognition \cite{snyder2018spoken,pappagari2020x}.


Most of the recent network architectures used for speaker embedding extraction are composed of a front-end feature extractor, a pooling layer, and a set of \gls{fc} layers. Lately, there have been several architectures proposed to encode audio utterances into speaker embeddings for different choices of network inputs, such as \cite{snyder2016deep,snyder2017deep,Chung18b,bhattacharya2019deep,safari2020self}. Using \gls{mfcc} features, \gls{tdnn} \cite{snyder2016deep, snyder2017deep} is the most currently used architecture. \gls{tdnn} is the x-vector front-end and consists of a stack of 1-D dilated \glspl{cnn}. The idea behind the use of \glspl{tdnn} is to encode a sequence of \gls{mfcc} into a more discriminative sequence of vectors by capturing long-term feature relations. 2-D \glspl{cnn} have also shown competitive results for speaker verification. There are Computer Vision architectures such as VGG \cite{Chung18b, chung2018voxceleb2, safari2019self} and ResNet \cite{bhattacharya2019deep, zhou2019deep, hajavi2019deep} that have been adapted to capture speaker discriminative information from the Mel-Spectrograms. In fact, Resnet34 has shown a better performance than \gls{tdnn} in the most recent speaker verification challenges \cite{chung2019voxsrc, zeinali2019but}. Finally, there are also some other attempts to work directly on the raw signal instead of using hand-crafted features \cite{ravanelli2018speaker, jung2018avoiding, jung2019rawnet}.

Given the encoded sequence from the front-end, a pooling layer is adopted to obtain an utterance-level representation. During the last few years, there are several studies addressing different types of pooling strategies \cite{cai2018exploring, xie2019utterance, jung2019spatial}. X-vector originally uses statistical pooling \cite{snyder2017deep}. Self attention mechanisms have been used to improve statistical pooling, such as \cite{zhu2018self}. In works like \cite{okabe2018attentive}, attention is used to extract better order features statistics. A wide set of pooling layers based on self attention have been proposed improving this vanilla self attention mechanism. In \cite{zhu2018self} several attentions are applied over the same encoded sequence, producing multiple context vectors. In our previous work \cite{safari2019self}, the encoded sequence is split into different heads and a different attention model is applied over each head sub-sequence. Non self attention mechanisms have also been proposed like \cite{li2020text}, where a mutual attention network is fed with the pair of utterances aimed to compare. 


In this paper we present a Double \gls{mha} pooling layer for speaker verification. 
The use of this layer is inspired by \cite{chen20182}, where Double \gls{mha} is presented as a double self attention block which captures feature statistics and makes adaptive feature
assignment over images. In this work this mechanism is used as a combination of two self attention pooling layers to create utterance-level speaker embeddings. Given a sequence of encoded representations from a \gls{cnn}, Self \gls{mha} first concatenates the context vector from $K$ head attentions applied over a $K$ sub-embedding sequences. An additional self attention mechanism is then applied over the multi-head context vector. This attention based pooling summarizes the set of head context vectors into a global speaker representation. This representation is pooled through a weighted average of the head context vectors, where the head weights are produced with the self attention mechanism. On the one hand, this approach allows the model to attend to different parts of the sequence, capturing at the same time different subsets of encoded representations. On the other hand, the pooling layer allows to select which head context vectors are the most relevant to produce the global context vector. In comparison with \cite{chen20182}, the second pooling layer operates over the head context vectors produced by a \gls{mha} instead of the global descriptors produced by a self multi attention mechanism applied over an image.


\section{Proposed Architecture}
\label{sec:DoubleMHA}

\begin{figure}[!t]
\centering
\includegraphics[width=3.2in]{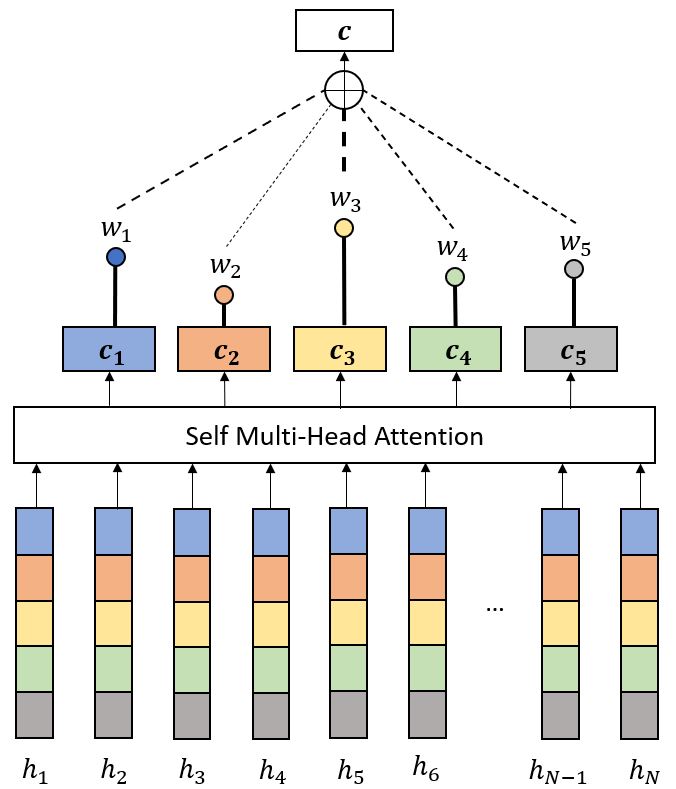}
\caption{An example of Double \gls{mha} Pooling with $5$ heads.}
\label{FIG: multihead-attn}
\end{figure}

Our proposed system architecture is illustrated in Figure \ref{FIG: systemDescription}. It utilizes a \gls{cnn}-based front-end which takes in a set of variable length mel-spectrogram features and outputs a sequence of speaker representations. These speaker representations are further subject to a Double \gls{mha} pooling which is the main contribution of this work. The Double \gls{mha} layer comprises a Self \gls{mha} pooling and an additional Self Attention layer that summarizes the information of each head context vector into a unique speaker embedding. The combination of Self \gls{mha} pooling together with this Self Head Attention layer provides us with a deeper self-attention pooling mechanism (Figure \ref{FIG: multihead-attn}). The speaker embedding obtained from the pooling layer is sent through a set of \gls{fc} layers to predict the speaker posteriors. This network architecture is trained with \gls{ams} loss \cite{liu2019large} as a speaker classifier so as to have a speaker embedding extractor.
 
\subsection{Front-End Feature Extractor}
Our feature extractor network is a larger version of the adapted VGG proposed in \cite{safari2019self}.  This \gls{cnn} comprises four convolution blocks, each of which contains two concatenated convolutional layers followed by a max pooling with a $2\times2$ stride. Hence given a spectrogram of $N$ frames, the VGG performs a down-sampling reducing its output into a sequence of $N/16$ representations. The output of the VGG $h\in\mathbb{R}^{M\times N/16\times D'}$ is a set of $M$ feature maps with $N/16\times D'$ dimension. These feature maps are concatenated into a unique vector sequence. This reshaped sequence of hidden states can now be defined as $h \in \mathbb{R}^{N/16\times D}$, where $D=MD'$ corresponds to the hidden state dimension.

\subsection{Self Multi-Head Attention Pooling}
\label{sec: SelfMHA}
 
The sequence of hidden states output from the front-end feature extractor can be expressed as $h=[h_1 h_2  ... h_N]$ with $h_t\in \mathbb{R}^{D}$. If we consider a number of $K$ heads for the \gls{mha} pooling, now we can define the hidden state as $h_t=[h_{t1}h_{t2}...h_{tK}]$ where $h_{tj} \in \mathbb{R}^{D/K}$.
Hence each feature vector is split into a set of sub-feature vectors of size $D/K$. In the same way, we have also a trainable parameter $u=[u_1u_2...u_K]$ where $u_j  \in \mathbb{R}^{D/K}$. A self attention operation is then applied over each head of the encoded sequences. The weights of each head alignment are defined as:

\begin{equation}
\centering
w_{tj} = \frac{\exp{\left(\frac{h^{T}_{tj}u_j}{\sqrt{d_h}}\right)}}{\sum^{N}_{l=1}\exp{\left(\frac{h^{T}_{lj}u_j}{\sqrt{d_h}}\right)}}
\end{equation}

where $w_{tj}$ corresponds to the attention weight of the head $j$ on the step $t$ of the sequence and $d_h$ corresponds to hidden state dimension $D/K$. If each head corresponds to a subspace of the hidden state, the weight sequence of that head can be considered as a \gls{pdf} from that subspace features over the sequence. We then compute a new pooled representation for each head in the same way than vanilla self attention:

\begin{equation}
\centering
c_j = \sum^{N}_{t=1}h^{T}_{tj}w_{tj}
\end{equation}

where $c_j\in \mathbb{R}^{D/K}$ corresponds to the utterance level representation from head $j$. The final utterance level representation is then obtained with the concatenation of the utterance level vectors from all the heads $c=[c_1c_2...c_k]$. This method allows the network to extract different kinds of information over different regions of the network. 

\subsection{Double Multi-Head Attention}

\begin{figure}[!t]
\centering
\includegraphics[width=3.4in]{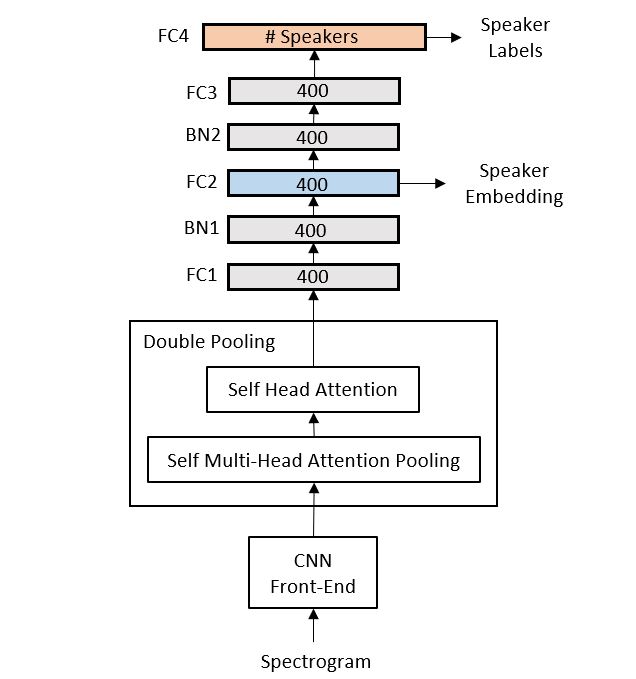}
\caption{System Architecture.}
\label{FIG: systemDescription}
\end{figure}

The main disadvantage of Self \gls{mha} pooling is that it assumes uniform head relevance. The output context vector is the concatenation of all head context vectors and it is used as input of the following dense layers. Double \gls{mha} does not assume that. Therefore, each utterance context vector is computed as a different linear combination of head context vectors. A summarized vector $c$ is then defined as a weighted average over the set of head context vectors $c_i$. Self attention is then used to pool the set of head context vectors $c_i$ and obtain an overall context vector $c$.



\begin{equation}
\centering
w'_{i} = \frac{\exp{(c^{T}_{i}u')}}{\sum^{K}_{l=1}\exp{(c^{T}_{l}u')}}
\end{equation}

\begin{equation}
\centering
c = \sum^{K}_{i=1}c^{T}_{i}w'_{i}
\end{equation}

where $w'_i$ corresponds to the aligned weight of each head and $u' \in \mathbb{R}^{D/K}$ is a trainable parameter. The context vector $c$ is then computed as the weighted average of the context vectors among heads. With this method, each utterance context vector is created scaling the information of the most/least relevant heads. Considering the whole pooling layer, Double \gls{mha} allows to capture different kind of speaker patterns in different regions of the input, and at the same time allows to weight the relevance of each of these patterns for each utterance.

 The number of heads used for this pooling defines both the context vector dimension and how the VGG feature maps are grouped. Considering the number of $M$ channels and $K$ heads, for each head we would create a $c_i$ context vector of $D'M/K$ dimension which contains a subset of $M/K$ feature maps. Therefore, as the number of heads grows larger, it allows Double \gls{mha} to consider more subsets of features while decreases the dimension of the final utterance-level context vector. This implies a trade-off between the number of features subsets we can create and how much compressed are these features in the context vector subspace.

\subsection{Fully-Connected Layers}
\label{sec: Dense-Layers}
The utterance-level speaker vector obtained from the pooling layer is fed into a set of four \gls{fc} layers (Figure \ref{FIG: systemDescription}). Each of the first two \gls{fc} layers is followed by a batch normalization \cite{ioffe2015batch} and \gls{relu} activations. A dense layer is adopted for the third \gls{fc} layer 
and the last \gls{fc} corresponds to the speaker classification layer. Since \gls{ams} is used to train the network, the third layer is set up without activation and batch normalization as proposed in \cite{liu2019large}. Once the network is trained, we can extract a speaker embedding from one of the intermediate \gls{fc} layers. According to \cite{liu2019large}, we consider the second layer as the speaker embedding instead of the third one. The output of this \gls{fc} layer then corresponds to the speaker representation that will be used for the speaker verification task.


\section{Experimental Setup}
\label{sec:setup}

The proposed system\footnote{Models are available at: \\  https://github.com/miquelindia90/DoubleAttentionSpeakerVerification} in this work has been assessed by VoxCeleb dataset \cite{Nagrani17, Chung18b}. VoxCeleb is a large multimedia database that contains more than one million 16kHz audio utterances for more than 6K celebrities. 
VoxCeleb has two different versions with several evaluation protocols. For our experiments, VoxCeleb2 development partition with no augmentation have been used to train both baseline and presented approaches.  
On the other hand, the performance of these systems have been evaluated with Vox1 test, Vox1-E, and Vox1-H conditions. These protocols contain sets of 37,611,  581,480, and 552,536 of Vox1 random pairs, respectively. Vox1 test only uses the test set, Vox1-E uses the whole development + test corpus and Vox1-H is restricted to audio pairs from same nationality and gender speakers. 

\begin{table}[!t]
\caption{\gls{cnn} Architecture. In and Out Dim. refers to the input and output feature maps of the layer. Feat Size refers to the dimension of each one of this output feature maps.}
\begin{center}
\begin{tabular}{|c|c|c|c|c|c|}
\hline
Layer & Size & In Dim. & Out Dim. & Stride & Feat Size\\ \hline
conv11 & 3x3 & 1 & 128 & 1x1 & Nx80\\ \hline
conv12 & 3x3 & 128 & 128 & 1x1 & Nx80\\ \hline
mpool1 & 2x2 & - & - & 2x2 & N/2x40\\ \hline
conv21 & 3x3 & 128 & 256 & 1x1 & N/2x40\\ \hline
conv22 & 3x3 & 256 & 256 & 1x1 & N/2x40\\ \hline
mpool2 & 2x2 & - & - & 2x2 & N/4x20\\ \hline
conv31 & 3x3 & 256 & 512 & 1x1 & N/4x20\\ \hline
conv32 & 3x3 & 512 & 512 & 1x1 & N/4x20\\ \hline
mpool3 & 2x2 & - & - & 2x2 & N/8x10\\ \hline
conv41 & 3x3 & 512 & 1024 & 1x1 & N/8x10\\ \hline
conv42 & 3x3 & 1024 & 1024 & 1x1 & N/8x10\\ \hline
mpool4 & 2x2 & - & - & 2x2 & N/16x5\\ \hline
 flatten & - & 1024 & 1 & - & N/16x5120\\ \hline
\end{tabular}
\end{center}

\label{TAB: Network Dimensions}
\end{table}

Two different baselines have been considered to compare with the presented approach. Double \gls{mha} pooling have been evaluated against two self attentive based  pooling methods: vanilla Self Attention and Self \gls{mha}. In order to evaluate them, these mechanisms have replaced the pooling layer of the system (Figure \ref{FIG: systemDescription}) without modifying any other block or parameter from the network. The speaker embeddings used for the verification tests have been extracted from the same \gls{fc} layer for each of the pooling methods.
Cosine distance has been used to compute the scores between pairs of speaker embeddings. 

The proposed network has been trained to classify variable-length speaker utterances. As input features we have used $80$ dimension log Mel Spectrograms with  $25$ms length Hamming windows and $10$ms window shift. 
The audio features have been normalized with \gls{cmn}. The \gls{cnn} encoder is then fed with $N\times 80$ spectrograms to obtain a sequence of $N/16\times 5120$ encoded hidden representations. For training we have used batches of $N=350$ frames audio chunks but for test the whole utterances have been encoded. The setup of the \gls{cnn} feature extractor can be found on Table \ref{TAB: Network Dimensions}. For the pooling layer we have tuned the number of heads for both Self \gls{mha} and Double \gls{mha}. For the presented \gls{cnn} setup we have considered $8$, $16$, and $32$ heads, which implies a head context vector $c_i$ of $640$, $320$, and $160$, respectively. Models with $64$ heads have been discarded due to the instability of their training. The last block of the system consists of four consecutive \gls{fc} layers. The first three dense layers have $400$ dimension. The last \gls{fc} layer has $5994$ dimension, which corresponds to the number of training speaker labels. Batch normalization has been applied only on the first two dense layers as mentioned in subsection 2.4. The network has been trained with \gls{ams} loss with $s=30$ and $m=0.4$ hyper-parameters. Batch size have been set to $128$ and Adam optimizer has been used to train all the models with a 1e-4 learning rate and a 1e-3 weight decay. Models have been trained for $100$ epochs using a learning rate annealing strategy. After each 15 epochs without validation improvement, learning rate was decayed by a $0.5$ factor. 

\vspace{-8pt}
\section{Results}

\begin{table*}[!t]
    \caption{Evaluation results on VoxCeleb 1 protocols. Head and global context vectors are referred to as $c_i$ and $c$.}
    \label{TAB: Results}
    \centering
    \vspace{4pt}
    \begin{tabular}{|l|ccc|cc|cc|cc|}
    \hline
    \multirow{2}{*}{\textbf{Approach}} & \multicolumn{3}{c|}{\textbf{Pooling Setup}}              & \multicolumn{2}{c|}{\textbf{Vox1 Test}} & \multicolumn{2}{c|}{\textbf{Vox1-E}} & \multicolumn{2}{c|}{\textbf{Vox1-H}} \\ \cline{2-10} 
    & \textbf{Heads} & \textbf{$c_i$ dimension} & \textbf{$c$ dimension} & \textbf{EER} & \textbf{DCF} & \textbf{EER} & \textbf{DCF}    & \textbf{EER}     & \textbf{DCF}    \\ \hline
    Attention & $1$ & $5120$ & $5120$ & $3.42$ & $0.0031$ & $3.42$ & $0.0029$ & $4.89$ & $0.0038$ \\ \hline
    \gls{mha} & $8$ & $640$ & $5120$ & $3.36$ & $0.0029$ & $3.44$ & $0.0029$ & $5.04$ & $0.004$ \\
    \gls{mha} & $16$ & $320$ & $5120$ & $3.43$ & $0.0032$ & $3.4$ & $0.0029$ & $4.9$ & $0.004$ \\
    \gls{mha} & $32$ & $160$ & $5120$ & $3.64$ & $0.0032$ & $3.68$ & 0.0031 & $5.35$ & $0.0042$ \\ \hline
    Double \gls{mha} & $8$ & $640$ & $640$ & $3.27$ & $0.0028$ & $3.23$ & $0.0028$ & $4.69$ & $0.0037$ \\
    Double \gls{mha} & $16$ & $320$ & $320$ & \textbf{3.19} & \textbf{0.0027} & $3.22$ & $0.0027$ & $4.67$ & $0.0038$ \\
    Double \gls{mha} & $32$ & $160$ & $160$ & $3.23$ & $0.0028$ & \textbf{3.18} & \textbf{0.0026} & \textbf{4.61} & \textbf{0.0036} \\ \hline
    \end{tabular}
    
\end{table*}


The proposed approach has been evaluated against different attention methods in the VoxCeleb text-independent speaker verification task. Performance is evaluated using \gls{eer} and \gls{dcf} calculated using $C_{FA}=1$, $C_{M}=1$, and $P_{T}=0.01 $. 
The results of this task are presented in Table \ref{TAB: Results}. Besides the mentioned metrics, both head $c_i$ and $c$ context vectors dimensions are shown for each presented pooling approach. 

Self Attention pooling has shown very similar results compared to Self \gls{mha} approaches. In comparison to Self Attention, Self \gls{mha} has shown better results in Vox1-test and Vox1-E protocols
with 8 heads and 16 heads, respectively. The relative improvement of these approaches compared to Self Attention are $1.75\%$ in terms of EER in Vox1 Test for 8 heads and $0.58\%$ in terms of EER in Vox1-E for the 16 heads model. Otherwise, Self Attention showed the best baseline result in Vox1-H  with a $4.89\%$ EER and a 0.0038 DCF. Compared to the best Self \gls{mha} approach, Self attention has only shown a relative improvement of $0.58\%$ in terms of EER. 
This similarity in the results indicates that Self \gls{mha} has not led to a noticeable performance improvement compared to vanilla Self Attention pooling. Double \gls{mha} have shown better results for all head values compared with both Self Attention and Self \gls{mha} approaches. In average, the $32$ heads model has outperformed all the baseline systems considering both EER and DCF metrics. For Vox1 Test, Vox1-E and Vox1-H, $32$ heads Double \gls{mha} have shown a $11.26\%$, $13.58\%$ and $13.83\%$ of EER relative improvement in comparison to the $32$ heads Self \gls{mha} model, respectively. This performance increase has also been shown for $8$ and $16$ heads in average for all the protocols with a $5,23\%$ and $5,62\%$ improvement, respectively. Hence Double \gls{mha} has provided the best results and it has shown to be more effective than Self \gls{mha} for all the head values.

As the results have shown, best performance in Double \gls{mha} based models has been achieved with $16$ and $32$ heads. We have included head and global context vector dimensions in Table \ref{TAB: Results} in order to analyze the relation between the number of heads and the models performance. As it was discussed in subsection 2.3, $c_i$ in Self \gls{mha} and both $c_i$ and $c$ dimensions in Double \gls{mha} are inversely proportional to the number of heads. Therefore, there is a trade-off between the number of attentions used over the encoded sequence and the amount of speaker information each attention is able to capture. Worst performance with Double \gls{mha} was achieved with $8$ heads. This setup implies that both $c_i$ and $c$ dimensions are $640$. Current state-of-the-art speaker embeddings have a dimension range between $200$ and $1500$ approximately. This means that there is still some margin to reduce the $c$ dimension and increase the number of attentions used. Increasing the number of heads has led to a better verification performance. The best Double \gls{mha} model has $32$ heads, whose context vector dimension $c$ is $160$. Here the speaker information was encoded into a lower dimension representation and the pooling was allowed to attend to $32$ different sub-sets of \gls{cnn} channels. Models with a larger number of heads have also been considered. However these models could not have been trained due to the narrow $c$ dimension, which led us to unstable training. Therefore Double \gls{mha} can be considered as a regularized extension of Self \gls{mha} that works as a bottleneck layer. On the other hand, the head number selection is also related to the \gls{cnn} setup. The results indicate that \gls{cnn} output feature maps are more efficiently grouped in subsets of $M/K=32$ channels, which correspond to sub-sequences of $160$ dimension embeddings. Considering these sets of $32$ context vectors pooled in that layer, these representations are efficiently averaged with Double \gls{mha} into unique $160$ dimension utterance-level speaker representations.

\vspace{-8pt}

\section{Conclusion}

In this paper we have implemented a Double Multi-Head Attention mechanism to obtain speaker embeddings at utterance level by pooling short-term representations. The proposed pooling layer is composed of a Self Multi-Head Attention pooling and a Self Attention mechanism that summarizes the context vectors of each head into a unique speaker vector. This pooling layer has been tested in a neural network based on a \gls{cnn}. The \gls{cnn} maps the spectrograms into sequences of speaker vectors. These vectors are the inputs to the proposed pooling layer, whose output activations are then connected to a set of dense layers. The network is trained as a speaker classifier and a bottleneck layer from these fully connected layers is used as speaker embedding. We have evaluated this approach with other pooling methods for the text-independent speaker verification task using the speaker embeddings and applying the cosine distance. The presented approach has outperformed both vanilla Self Attention and Self Multi-Head Attention poolings.

\bibliographystyle{IEEEbib}
\bibliography{mybib}
\end{document}

%% file: glossary.tex
\newacronym{dl}{DL}{Deep Learning}
\newacronym{dbn}{DBN}{Deep Belief Network}
\newacronym{dnn}{DNN}{Deep Neural Network}
\newacronym{rbm}{RBM}{Restricted Boltzmann Machine}
\newacronym{cnn}{CNN}{Convolutional Neural Network}
\newacronym{rnn}{RNN}{Recurrent Neural Network}
\newacronym{lstm}{LSTM}{Long Short-Term Memory}
\newacronym{dbm}{DBM}{Deep Boltzmann Machine}
\newacronym{cd}{CD}{Contrastive Divergence}
\newacronym{urbm}{URBM}{Universal RBM}
\newacronym{udbn}{UDBN}{Universal DBN}
\newacronym{relu}{ReLU}{Rectified Linear Unit}
\newacronym{vrelu}{VReLU}{Variable ReLU}
\newacronym{ubm}{UBM}{Universal Background Model}
\newacronym{gmm}{GMM}{Gaussian Mixture Model}
\newacronym{map}{MAP}{Maximum a Posteriori}
\newacronym{nist}{NIST}{National Institute of Standard and Technology}
\newacronym{sre}{SRE}{Speaker Recognition Evaluation}
\newacronym{nap}{NAP}{Nuisance Attribute Projection}
\newacronym{jfa}{JFA}{Joint Factor Analysis}
\newacronym{lid}{LID}{Language Identification}
\newacronym{lda}{LDA}{Linear Discriminant Analysis}
\newacronym{vad}{VAD}{Voice Activity Detection}
\newacronym{plda}{PLDA}{Probabilistic Linear Discriminant Analysis}
\newacronym{wccn}{WCCN}{Within-Class Covariance Normalization}
\newacronym{asr}{ASR}{Automatic Speech Recognition}
\newacronym{det}{DET}{Detection Error Tradeoff}
\newacronym{eer}{EER}{Equal Error Rate}
\newacronym{dcf}{DCF}{Detection Cost Function}
\newacronym{mindcf}{minDCF}{minimum DCF}
\newacronym{far}{FAR}{False Acceptance Rate}
\newacronym{frr}{FRR}{False Rejection Rate}
\newacronym{fbe}{FBE}{Filter Bank Energy}
\newacronym{ff}{FF}{Frequency Filtering}
\newacronym{hmm}{HMM}{Hidden Markov Model}
\newacronym{lpc}{LPC}{Linear Predictive Coefficient}
\newacronym{lpr}{LPR}{Log Posterior Ratio}
\newacronym{mfcc}{MFCC}{Mel-Frequency Cepstral Coefficient}
\newacronym{lfcc}{LFCC}{Linear Frequency Cepstral Coefficient}
\newacronym{nn}{NN}{Neural Network}
\newacronym{pca}{PCA}{Principal Component Analysis}
\newacronym{plp}{PLP}{Perceptual Linear Predictive}
\newacronym{sdc}{SDC}{Shifted Delta Cepstrum}
\newacronym{psd}{PSD}{Power Spectral Density}
\newacronym{rbf}{RBF}{Radial Basis Function}
\newacronym{dct}{DCT}{Discrete Cosine Transform}
\newacronym{snr}{SNR}{Signal-to-Noise Ratio}
\newacronym{svm}{SVM}{Support Vector Machine}
\newacronym{cmn}{CMN}{Cepstral Mean Normalization}
\newacronym{cmvn}{CMVN}{Cepstral Mean and Variance Normalization}
\newacronym{cdf}{CDF}{Cumulative Distribution Function}
\newacronym{pdf}{pdf}{probability density function}
\newacronym{em}{EM}{Expectation-Maximization}
\newacronym{sv}{SV}{Speaker Verification}
\newacronym{sr}{SR}{Speaker Recognition}
\newacronym{an}{AN}{Adversarial Network}
\newacronym{en}{EN}{Encoding Network}
\newacronym{dn}{DN}{Discriminative Network}
\newacronym{mlp}{MLP}{Multilayer Perceptron}
\newacronym{gru}{GRU}{Gated Recurrent Unit}
\newacronym{tdnn}{TDNN}{Time Delay Neural Network}
\newacronym{nmt}{NMT}{Neural Machine Translation}
\newacronym{fc}{FC}{Fully Connected}
\newacronym{mha}{MHA}{Multi-Head Attention}
\newacronym{ams}{AMS}{Additive Margin Softmax}
\newacronym{tddn}{TDNN}{Time Delay Neural Network}